\documentclass[conference]{IEEEtran}
\IEEEoverridecommandlockouts

\usepackage[numbers,sort&compress]{natbib}
\usepackage{amsmath,amssymb,amsfonts}
\usepackage{algorithmic}
\usepackage{graphicx}
\usepackage{textcomp}
\usepackage{xcolor}
\usepackage{graphicx}
\usepackage{float}
\usepackage{booktabs}
\usepackage{multirow}
\usepackage{threeparttable}
\usepackage{afterpage}
\usepackage{stfloats}
\usepackage{graphicx}
\usepackage{amsmath}
\usepackage{adjustbox}
\usepackage{algorithm}
\usepackage{algorithmic}
\usepackage{threeparttable}

\usepackage{enumitem}
\usepackage{amsmath}

\DeclareRobustCommand*{\IEEEauthorrefmark}[1]{%
	\raisebox{0pt}[0pt][0pt]{\textsuperscript{\footnotesize\ensuremath{#1}}}}

\def\BibTeX{{\rm B\kern-.05em{\sc i\kern-.025em b}\kern-.08em
    T\kern-.1667em\lower.7ex\hbox{E}\kern-.125emX}}
\begin{document}

\title{SegSEM: Enabling and Enhancing SAM2 for SEM Contour Extraction}


\author{
	\IEEEauthorblockN{
		Da Chen\IEEEauthorrefmark{1*}, Guangyu Hu\IEEEauthorrefmark{2*}, Kaihong Xu\IEEEauthorrefmark{1}, Kaichao Liang\IEEEauthorrefmark{1}, Songjiang Li\IEEEauthorrefmark{1}, \\ Wei Yang\IEEEauthorrefmark{3},
		XiangYu Wen\IEEEauthorrefmark{4}, Mingxuan Yuan\IEEEauthorrefmark{1,2}
	}
	\IEEEauthorblockA{\IEEEauthorrefmark{1}Noah's Ark Lab, Huawei Technologies, China \quad
		\IEEEauthorrefmark{2}HKUST, Hong Kong}
	\IEEEauthorblockA{\IEEEauthorrefmark{3}HiSilicon, Huawei Technologies, China \quad
		\IEEEauthorrefmark{4}CURE Lab, CUHK, Hong Kong}
	\IEEEauthorblockA{915102610106@njust.edu.cn, ghuae@connect.ust.hk,\\ \{xukaihong1, liangkaichao, songjiang.li, Yuan.Mingxuan\}@huawei.com, \\ yangwei59@hisilicon.com, 1155186676@link.cuhk.edu.hk}
	\IEEEauthorblockA{\IEEEauthorrefmark{*}Both authors contributed equally to this research.}
}
\maketitle
\vspace{-1cm}

\begin{abstract}
Extracting high-fidelity 2D contours from Scanning Electron Microscope (SEM) images is critical for calibrating Optical Proximity Correction (OPC) models. While foundation models like Segment Anything 2 (SAM2) are promising, adapting them to specialized domains with scarce annotated data is a major challenge. This paper presents a case study on adapting SAM2 for SEM contour extraction in a few-shot setting. We propose \textbf{\texttt{SegSEM}}, a framework built on two principles: a data-efficient fine-tuning strategy that adapts by selectively training only the model's encoders, and a robust hybrid architecture integrating a traditional algorithm as a confidence-aware fallback. Using a small dataset of 60 production images, our experiments demonstrate this methodology's viability. The primary contribution is a methodology for leveraging foundation models in data-constrained industrial applications.
\end{abstract}


\section{Introduction}
\begin{figure}[htbp]
	\centering
	\includegraphics[width=0.7\linewidth]{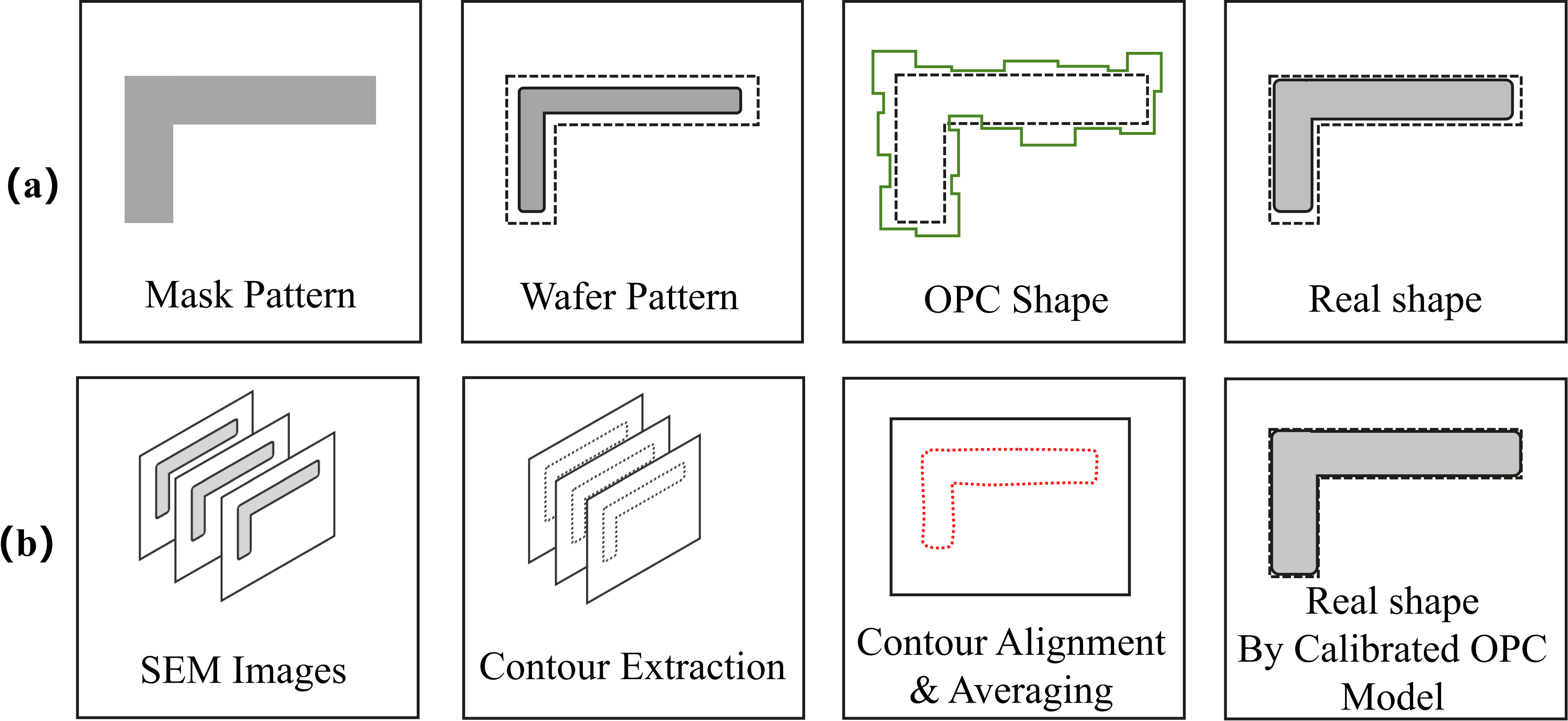}
	\caption{(a) Optical Proximity Correction (OPC) standard procedure; (b) using SEM image to calibrate OPC model}
	\label{opc-procedure}
\end{figure}

In advanced semiconductor manufacturing, ensuring the fidelity of circuit patterns on the wafer is paramount. Optical Proximity Correction (OPC) is a cornerstone technology that pre-distorts photomask patterns to compensate for optical and process effects during photolithography, as shown in Fig.~\ref{opc-procedure}(a). The accuracy of OPC models, which dictate these corrections, depends heavily on high-quality on-wafer measurement data for calibration and verification. While traditional methods rely on one-dimensional Critical Dimension (CD) measurements from Scanning Electron Microscopes (SEMs), these are insufficient for capturing the complex two-dimensional nature of modern circuit layouts. Consequently, the direct extraction of 2D contours from SEM images, depicted in Fig.~\ref{opc-procedure}(b), has become indispensable for building accurate OPC models and ensuring manufacturing precision \cite{taberySEMImageContouring2007a,vasekSEMcontourbasedOPCModel2007,weisbuchAssessingSEMContour2014,zhengOptimalOPCModel2022,weisbuchCalibratingEtchModel2015,lesireAcceleratingProcessRobustness2024}.

\setlength{\textfloatsep}{8pt plus 0.5pt minus 0.5pt}

\begin{figure}[htbp]
	\centering
	\includegraphics[width=0.7\linewidth]{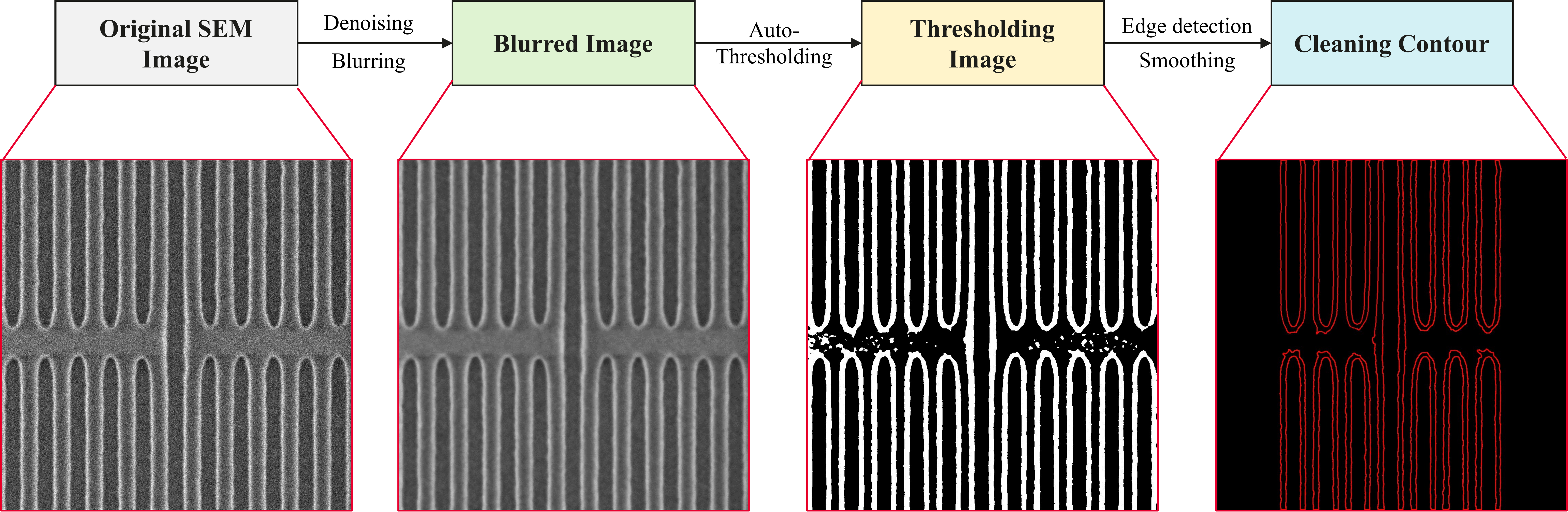}
	\caption{The overview of traditional SEM contour extraction method}
	\label{overall-traditional-structure}
\end{figure}

However, extracting high-fidelity contours from SEM images is challenging due to low signal-to-noise ratios, edge ambiguity, and significant process-induced variations. Traditional image processing pipelines (Fig.~\ref{overall-traditional-structure}), often based on thresholding, struggle with this variability, while standard deep learning models require large annotated datasets that are prohibitively expensive in this domain \cite{zhao2024adaptive,pang2024suppression}. This creates a dilemma: \textit{\textbf{how to achieve robust extraction without massive labeled datasets?}}
\par The advent of foundation models like Segment Anything Model (SAM) offers a path forward, given their powerful few-shot generalization capabilities \cite{gainesScanningElectronMicroscope2024,kousakaAutomatedCellStructure2025}. Yet, their direct application to a specialized domain like SEM imaging yields suboptimal results. This frames our core research question: \textit{What is an effective and data-efficient strategy for adapting a foundation model to a high-precision industrial task?}

In this paper, we conduct a case study to answer this question by focusing on SEM contour extraction. We propose \textbf{\texttt{SegSEM}}, a hybrid framework designed to explore the feasible adaptation of the state-of-the-art \texttt{SAM2} model in a few-shot industrial setting. Our work does not aim to claim state-of-the-art performance, but rather to contribute a practical methodology for leveraging foundation models where data is scarce.

Our core contributions are three-fold, focusing on the adaptation methodology:
\begin{enumerate}[leftmargin=*,noitemsep,topsep=0pt]
    \item \textbf{A Data-Efficient Adaptation Strategy}: We demonstrate that by freezing the pre-trained mask decoder and fine-tuning only the vision encoders, SAM2 can be effectively adapted to the SEM domain using a minimal dataset (e.g., 60 annotated images). This finding provides a valuable, resource-efficient blueprint for domain-specific adaptation of vision foundation models.
    \item \textbf{A Robust Hybrid Architecture Design}: We propose \textbf{\texttt{SegSEM}}, a framework that integrates the fine-tuned SAM2 with a traditional image processing algorithm via an intelligent, confidence-based fallback mechanism. This design choice acknowledges the limitations of any single model in production environments and prioritizes system-level reliability.
    \item \textbf{An Empirical Analysis of SAM2 in the SEM Domain}: We provide a detailed analysis of the adapted SAM2's performance under varying process conditions (i.e., exposure levels), identifying its strengths and failure modes. This empirical study contributes to a deeper understanding of how foundation models behave when transferred to high-precision industrial applications.
\end{enumerate}


\section{Related Work}
\label{sec:related}
Existing SEM contour extraction methods fall into two main categories. Traditional image processing pipelines, relying on filters and thresholding, are fast but often fail under the noise and process variations common in production environments \cite{heCDSEMContourExtraction2022, zhangMulticomponentSegmentationXray2017}. Supervised deep learning methods, such as CNN-based architectures (\texttt{Mask R-CNN}, \texttt{YOLO}), offer higher accuracy but demand large annotated datasets, which are prohibitively expensive to acquire in this domain \cite{liaoApplyingMaskRCNN2024, kangASFYOLONovelYOLO2024}.
\par More recently, foundation models like the Segment Anything Model (\texttt{SAM}) \cite{raviSAM2Segment} have shown promise due to their few-shot generalization capabilities \cite{shiSegmentAnythingModel2025, gainesScanningElectronMicroscope2024}. However, their direct application to specialized domains like SEM imaging yields suboptimal performance because they are not tuned for the unique noise and texture profiles. A critical gap thus exists in developing a data-efficient strategy to adapt these powerful models for high-precision industrial tasks. Our work addresses this gap by proposing a systematic framework for adapting \texttt{SAM2} and integrating it into a robust, production-ready workflow.

\section{Methodology}
\label{sec:method}
\setlength{\dbltextfloatsep}{5pt plus 1pt minus 1pt}  
\begin{figure*}[htbp]
	\centering
	\includegraphics[width=0.7\linewidth]{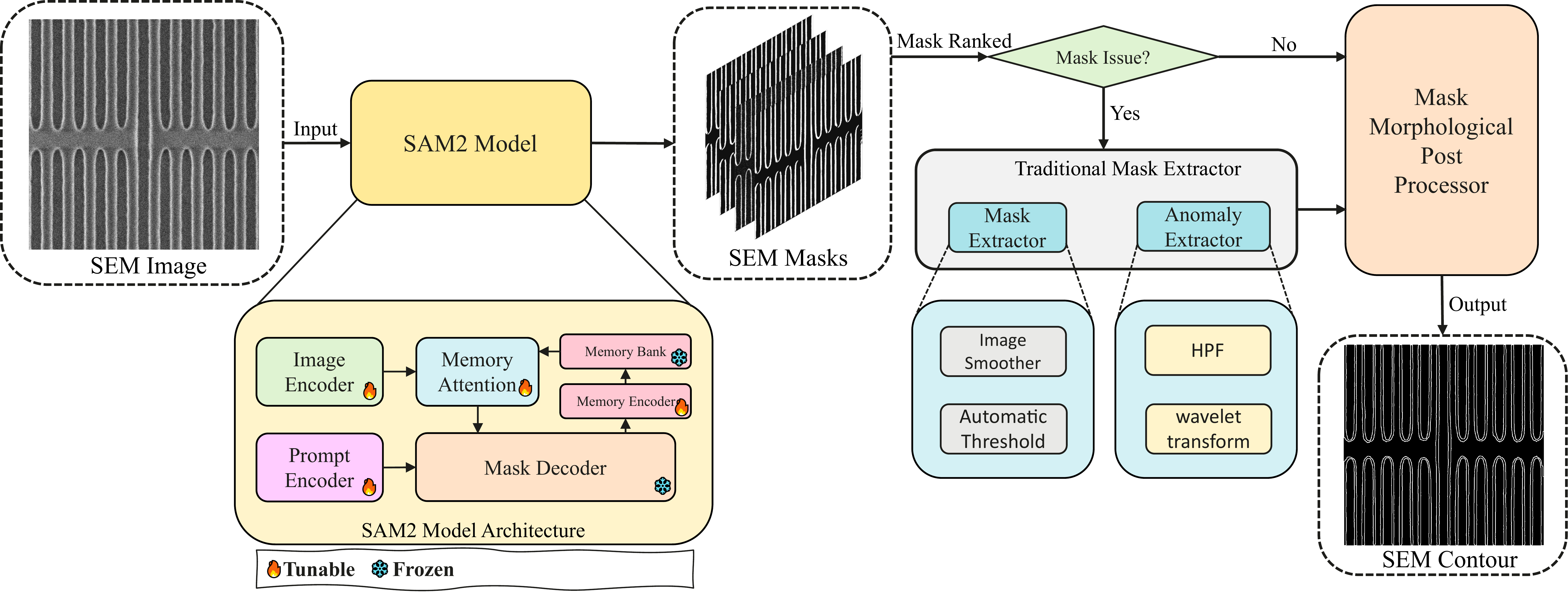}
	\caption{The overview of SEM contour extraction method based on \texttt{SAM2} network and enhanced mask extractor (Left: Fine-tuned \texttt{SAM2} model with unfrozen layers. Right: Automatic algorithm when \texttt{SAM2} fails to extract an effective mask.)}
	\label{overall-structure}
\end{figure*}

The proposed \texttt{SegSEM} framework is designed as a robust, hybrid system that leverages the strengths of both foundation models and traditional computer vision techniques. This section details its three core components: the overall architecture, the data-efficient adaptation strategy for \texttt{SAM2}, and the reliability-ensuring fallback module.

\subsection{The \texttt{SegSEM} Framework}
As depicted in Fig.~\ref{overall-structure}, the \texttt{SegSEM} pipeline begins with the fine-tuned \texttt{SAM2} model, which processes a raw SEM image to generate multiple mask candidates. Each candidate is assigned a confidence score based on model-inferred quality metrics. The framework automatically selects the highest-scoring mask that satisfies a set of predefined quality criteria. This selected mask then undergoes a post-processing step involving morphological operations to refine its boundaries before final contour extraction.

Crucially, if no candidate mask from \texttt{SAM2} meets the quality threshold, the framework seamlessly activates its auxiliary fallback module. This dual-path design ensures that a valid contour is always produced, making the system robust enough for deployment in a production environment where failure is not an option.

\subsection{Core Component: Data-Efficient Adaptation of \texttt{SAM2}}
A key challenge in applying foundation models to specialized domains is overcoming the domain gap without incurring massive annotation costs. Our approach addresses this through a targeted, data-efficient adaptation strategy for \texttt{SAM2}, which consists of a selective fine-tuning method and a custom loss function.

\subsubsection{Selective Fine-Tuning Strategy}
Our training dataset comprises only 60 SEM-mask pairs, covering a range of exposure conditions. To maximize learning from this limited data, we adopt a specific fine-tuning strategy. We hypothesize that the pre-trained mask decoder of \texttt{SAM2} already possesses a strong generalizable ability to generate high-quality mask structures. In contrast, the image and prompt encoders, having never been exposed to SEM-specific noise and texture patterns, are the primary sources of domain gap.

\begin{figure}[htbp]
	\centering
	\includegraphics[width=0.7\linewidth]{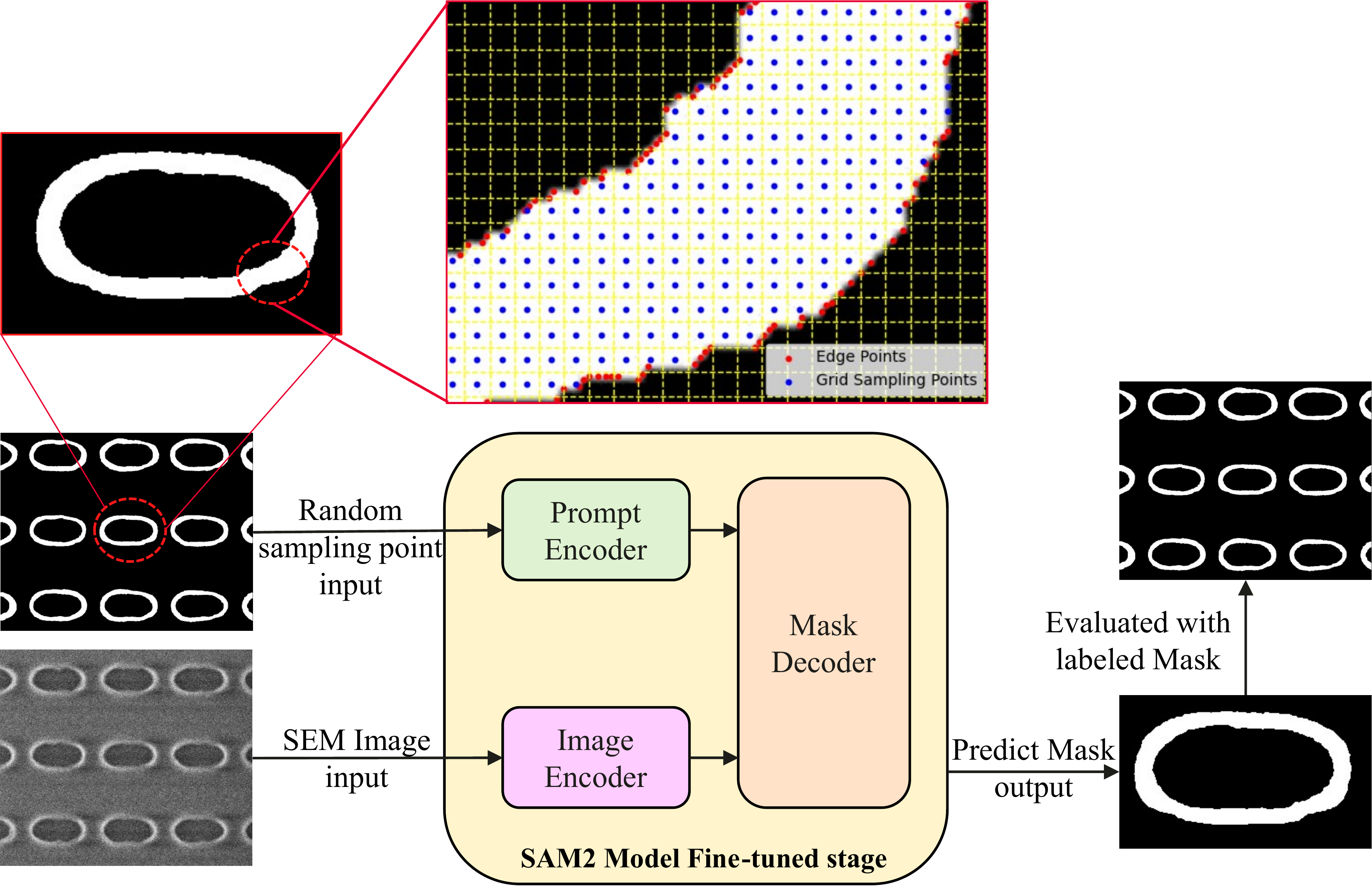}
	\caption{Training process of the \texttt{SAM2} model}
	\label{train-sam2-process}
\end{figure}

Therefore, we freeze the weights of the mask decoder and exclusively fine-tune the image and prompt encoders. This approach forces the encoders to learn representations tailored to SEM imagery while preventing the decoder from overfitting to the small dataset, thereby preserving its robust generalization capabilities. As shown in Fig.~\ref{train-sam2-process}, training is performed using raw SEM images paired with ground-truth masks. For each training instance, a single, randomly sampled point from within the ground-truth mask is provided as a prompt, which serves as a form of implicit data augmentation to further enhance robustness.

\subsubsection{Custom Loss Function for High-Fidelity Masks}
To guide the fine-tuning process toward generating geometrically precise and structurally sound masks, we design a multi-component loss function, shown in Equation~\ref{sam2_loss}. It augments the standard segmentation loss ($\mathcal{L}_{\text{seg}}$) with three task-specific terms:
\begin{enumerate}[leftmargin=*,noitemsep,topsep=0pt]
    \item \textbf{Mask IoU Reward}, which encourages a higher overlap between the predicted mask and the ground truth.
    \item \textbf{Mask Count Penalty}, which discourages the generation of multiple spurious mask fragments.
    \item \textbf{Edge Placement Error (EPE) Penalty} \cite{zhouEvaluationHotspotsEPE2024}, which specifically penalizes deviations at the mask boundary, crucial for CD measurement accuracy.
\end{enumerate}

\begin{equation}
	\begin{aligned}
		\mathcal{L}_{\text{\texttt{SAM2}}} = \mathcal{L}_{\text{seg}}
		& + \lambda_{\text{IoU}} \cdot \mathcal{R}_{\text{IoU}} \\
        & - \lambda_{\text{count}} \cdot \mathcal{P}_{\text{count}}
		 - \lambda_{\text{EPE}} \cdot \mathcal{P}_{\text{EPE}}
	\label{sam2_loss}
	\end{aligned}
\end{equation}
Here, $\mathcal{R}$ denotes a reward term, and $\mathcal{P}$ denotes a penalty term. The weights ($\lambda$) are determined empirically to balance the contribution of each component to the total gradient.

\setlength{\textfloatsep}{8pt plus 1pt minus 1pt}

This per-step stochastic point sampling acts as an implicit augmentation that mitigates overfitting and provides spatially localized, robust guidance to \texttt{SAM2} during training, thereby improving its generalization across noise and exposure variations.

Additionally, we directly input raw SEM images into the image encoder without applying blurring or denoising, as such preprocessing tends to distort the inherent random noise patterns. These noise features are essential for accurate segmentation, and altering them impairs the \texttt{SAM2} model’s ability to learn representative characteristics, ultimately reducing mask accuracy.

\subsection{Ensuring Reliability: The Hybrid Architecture with Fallback Module}
\begin{figure}[htbp]
  \centering
  \includegraphics[width=0.7\linewidth]{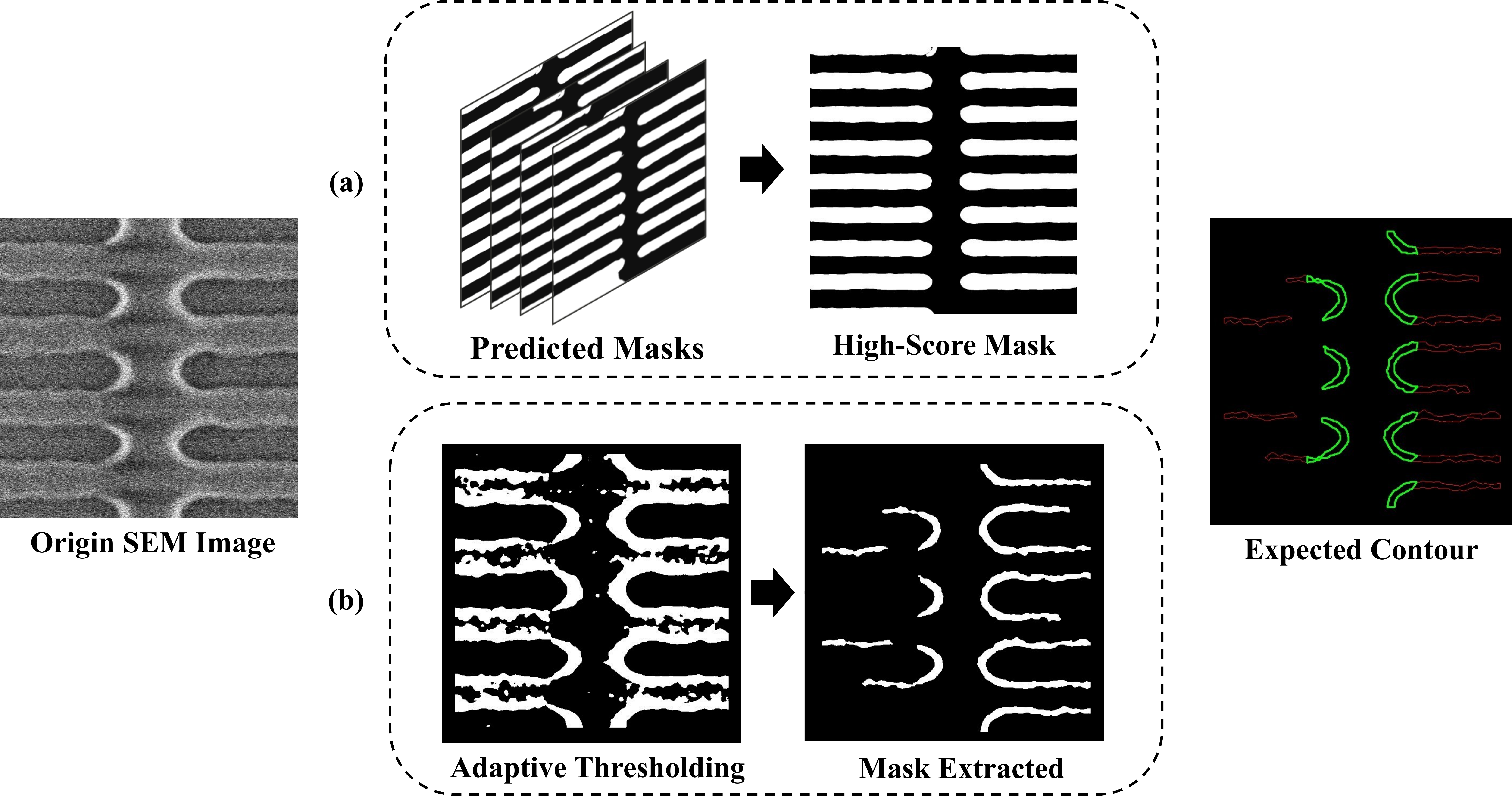}
  \caption{Example of fallback activation: (a) Failure of \texttt{SAM2} alone; (b) Successful extraction by the fallback module.}
  \label{fig:dual_failure}
\end{figure}

A core design principle of \texttt{SegSEM} is ensuring system-level reliability. While the fine-tuned \texttt{SAM2} performs well on a wide range of images, it can still produce flawed masks (e.g., disjointed fragments or geometrically incorrect shapes, see Fig.~\ref{fig:dual_failure}a) under extreme noise or exposure conditions. To mitigate this, we designed a hybrid architecture that incorporates a traditional algorithm as a confidence-aware fallback path.

\subsubsection{Automated Mask Quality Assessment and Fallback}
The decision to use the fallback path is governed by a deterministic, three-stage quality check on the \texttt{SAM2} mask candidate, $M_{\mathrm{SAM}}$:
\begin{enumerate}[leftmargin=*,noitemsep,topsep=0pt]
    \item \textbf{Confidence Check}: The model's predicted IoU score must exceed a threshold $\tau_{\text{conf}}$ (e.g., 0.90).
    \item \textbf{Topological Check}: A connected-component analysis ensures the mask is a single, contiguous region.
    \item \textbf{Geometric Check}: The mask's area and aspect ratio must fall within a plausible range derived from the design layout.
\end{enumerate}
If $M_{\mathrm{SAM}}$ fails any check, it is discarded, and the fallback path is activated. The fallback module then processes the image using a robust traditional algorithm (Sauvola's adaptive thresholding \cite{sauvola2000adaptive}) to generate a reliable mask, $M_{\mathrm{fallback}}$. This ensures a valid output is always produced.

\begin{figure*}[htbp]
	\centering
	\includegraphics[width=0.7\linewidth]{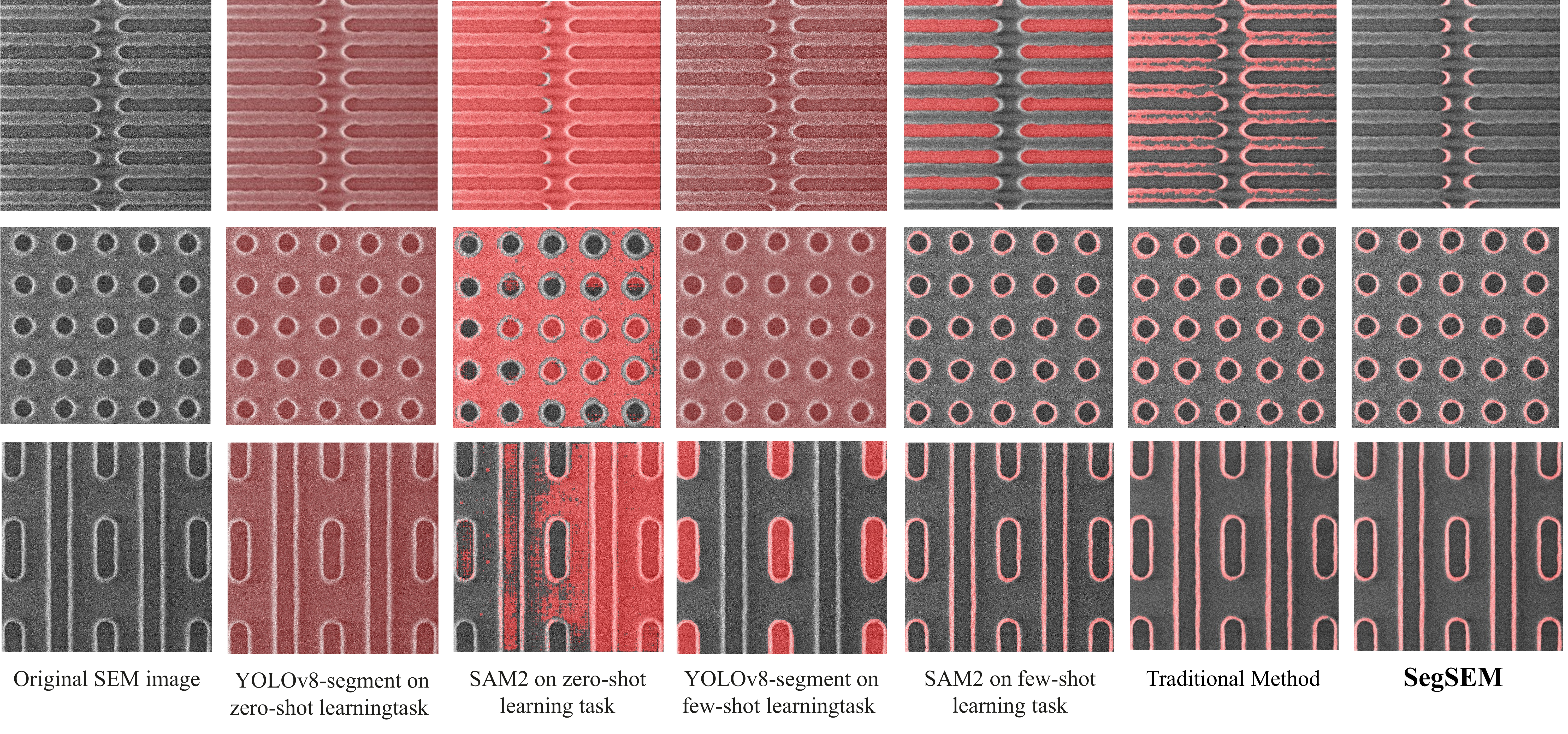}
	\caption{Comparison of processing results for YOLO and \texttt{SAM2} models in zero-shot and few-shot tasks, alongside the standard and proposed workflows for SEM contour extraction}
	\label{different-performance}
\end{figure*}

\subsection{Final Contour Generation}
Regardless of its origin (\texttt{SAM2} or fallback), the final accepted mask, $M_{\mathrm{final}}$, undergoes a concluding refinement stage. This stage employs a sequence of morphological operations (e.g., closing and hole filling) to eliminate minor imperfections such as small holes or jagged edges. Finally, the clean, high-fidelity SEM contour is extracted from the refined binary mask using the Sobel edge detection operator \cite{raniEdgeDetectionScanning2020}.

\section{Experiments and Results}
\label{sec:experiments}
\subsection{Experimental Setup and Results}

\noindent\textbf{Dataset and Setup.} We use a challenging dataset of 60 production SEM images with ground-truth masks, covering diverse exposure conditions. We partition it into 50 for training and 10 for testing (augmented to 50 samples per case). Models were trained on an Ascend 910B GPU. Our \texttt{SAM2-L} was fine-tuned with an AdamW optimizer, a $1\times10^{-5}$ learning rate, and a crucial early stopping mechanism that halted training in under 1,000 iterations, preventing overfitting. We compare \texttt{SegSEM} against three baselines: a \textbf{Traditional Method} (Sauvola's), a \textbf{\texttt{SAM2}-Only} version, and a fine-tuned \textbf{\texttt{YOLOv8n-seg}} model. Performance is measured by Intersection over Union (IoU) and F1 Score.

\begin{table*}[htbp]
	\caption{Quantitative Comparison of \texttt{SegSEM} and the Standard Baseline Across Diverse Exposure Conditions}
	\centering
	\label{different-exposure-results}
	\belowrulesep=0pt
	\aboverulesep=0pt
	\renewcommand{\arraystretch}{1.5}  
	\resizebox{0.9\linewidth}{!}{
		\begin{threeparttable}
			\begin{tabular}{c|c c| c c c c | c | c} 
				\toprule[0.7mm]
				Exposure Case & Average Exposure\tnote{1} & Number of Samples\tnote{2} & IoU\tnote{3} (Standard\tnote{*} / \textbf{\texttt{SAM2}\tnote{**}}\quad) & Precision\tnote{3} (Standard\tnote{*} / \textbf{\texttt{SAM2}\tnote{**}}\quad) & Recall\tnote{3} (Standard\tnote{*} / \textbf{\texttt{SAM2}\tnote{**}}\quad) & F1 Score\tnote{3} (Standard\tnote{*} / \textbf{\texttt{SAM2}\tnote{**}}\quad) &
                Execution Time\tnote{4} & Fallback Used\tnote{5}\\
				\midrule
				Case 1 & 10876 & 50 & 0.827 ±0.074 / \textbf{0.876 ±0.016} & 0.858 ±0.037 / \textbf{0.902 ±0.012} & 0.878 ±0.094 / \textbf{0.965 ±0.022} & 0.876 ±0.049 / \textbf{0.938 ±0.015} & \textbf{6463.59} & 4 \\
				\midrule
				Case 2 & 11487 & 50 & 0.753 ±0.092 / \textbf{0.897 ±0.013} & 0.892 ±0.032 / \textbf{0.965 ±0.009} & 0.828 ±0.064 / \textbf{0.927 ±0.013} & 0.859 ±0.069 / \textbf{0.946 ±0.011} & \textbf{6154.63} & 1\\
				\midrule
				Case 3 & 11987 & 50 & 0.854 ±0.098 / \textbf{0.883 ±0.011} & 0.882 ±0.057 / \textbf{0.894 ±0.032} & 0.964 ±0.074 / \textbf{0.987 ±0.017} & 0.921 ±0.073 / \textbf{0.938 ±0.011} & \textbf{6312.67} & 6\\
				\midrule
				Case 4 & 12089 & 50 & 0.815 ±0.077 / \textbf{0.897 ±0.013} & 0.921 ±0.044 / \textbf{0.930 ±0.013} & 0.876 ±0.074 / \textbf{0.962 ±0.005} & 0.898 ±0.089 / \textbf{0.946 ±0.021} & \textbf{6237.87} & 3\\
				\midrule
				Case 5 & 12108 & 50 & 0.858 ±0.051 / \textbf{0.873 ±0.013} & 0.887 ±0.066 / \textbf{0.889 ±0.015} & 0.963 ±0.098 / \textbf{0.990 ±0.001} & 0.923 ±0.069 / \textbf{0.932 ±0.023} & \textbf{5901.94} & 5\\
				\midrule
				Case 6 & 12651 & 50 & 0.832 ±0.091 / \textbf{0.882 ±0.014} & 0.911 ±0.036 / \textbf{0.929 ±0.006} & 0.906 ±0.089 / \textbf{0.979 ±0.014} & 0.908 ±0.079 / \textbf{0.937 ±0.015} & \textbf{6022.95} & 4\\
				\midrule
				Case 7 & 14152 & 50 & 0.705 ±0.083 / \textbf{0.906 ±0.013} & 0.819 ±0.058 / \textbf{0.943 ±0.011} & 0.836 ±0.092 / \textbf{0.958 ±0.012} & 0.827 ±0.059 / \textbf{0.951 ±0.013} & \textbf{5925.17} & 3\\
				\midrule
				Case 8 & 14736 & 50 & 0.620 ±0.083 / \textbf{0.912 ±0.006} & 0.931 ±0.056 / \textbf{0.953 ±0.007} & 0.649 ±0.061 / \textbf{0.955 ±0.012} & 0.765 ±0.099 / \textbf{0.954 ±0.017} & \textbf{6009.53} & 2\\
				\midrule
				Case 9 & 15132 & 50 & 0.780 ±0.075 / \textbf{0.893 ±0.011} & 0.875 ±0.077 / \textbf{0.902 ±0.013} & 0.878 ±0.094 / \textbf{0.989 ±0.005} & 0.877 ±0.053 / \textbf{0.944 ±0.014} & \textbf{6142.77} & 2\\
				\midrule
				Case 10 & 20038 & 50 & 0.773 ±0.094 / \textbf{0.820 ±0.014} & 0.831 ±0.076 / \textbf{0.867 ±0.011} & 0.917 ±0.076 / \textbf{0.938 ±0.008} & 0.872 ±0.086 / \textbf{0.901 ±0.012} & \textbf{6238.92} & 6 \\
				\midrule
				\textbf{Overall Improvement} & - & - & \textbf{13.07\% $\uparrow$} & \textbf{4.17\% $\uparrow$} & \textbf{10.98\% $\uparrow$} & \textbf{7.58\% $\uparrow$} & \textbf{6141.00} & - \\
				\bottomrule[0.5mm]
			\end{tabular}
			\begin{tablenotes}
				\footnotesize
				\item[1] The average exposure was computed using the Laplacian operator.
				\item[2] The number of sample SEM images used in the experiment.
				\item[3] The IoU (Intersection over Union), Precision, Recall, and F1 Score metrics were computed as the averages over the sample images.
                \item[4] Execution Time (ms) denotes the average time per batch measured on an Ascend 910B accelerator.
                \item[5] Denotes the number of test samples in which the SAM2 output did not meet the quality criteria and the fallback (traditional) method was consequently applied.
				\item[*] Standard SEM contour extraction process using traditional algorithm.
				\item[**] Our SAM2-based SEM contour extraction process fine-tuned on various exposure scenarios.
			\end{tablenotes}
		\end{threeparttable}
	}
\end{table*}

\noindent\textbf{Analysis of the Hybrid Framework.}
Table~\ref{different-exposure-results} shows that \texttt{SegSEM} consistently outperforms the traditional baseline. It achieves an average IoU of 0.884 (\textbf{13.07\% relative improvement}) and an F1 score of 0.941 (\textbf{7.58\% relative improvement}), with a significantly lower standard deviation, indicating enhanced stability. The ``Fallback Used'' column shows the fallback was triggered in only 36 of 500 cases (7.2\%), primarily in extreme exposure scenarios. This confirms the fine-tuned \texttt{SAM2} is highly effective (handling \>92\% of cases) while the fallback is crucial for system-level robustness.

\noindent\textbf{Ablation Study of the Adaptation Strategy.}
To validate our selective fine-tuning strategy, we compared three configurations: (1) Finetune Encoders Only (Our Method); (2) Finetune Decoder Only; and (3) Finetune All. The results (omitted for brevity, see supplementary) unequivocally support our strategy. Fine-tuning only the encoders successfully adapted the model by learning SEM-specific features while preserving the decoder's powerful generalization, achieving the best performance. Fine-tuning the decoder only was insufficient for domain adaptation, and fine-tuning all components led to overfitting.


\noindent\textbf{Discussion and Limitations.}
Our results confirm the viability of adapting foundation models for specialized, few-shot industrial tasks. However, we acknowledge this study's limitations: we did not evaluate the impact on downstream OPC tasks. These aspects are critical for production and represent important future work. This paper serves as a foundational case study in an industrial environment, providing a methodological blueprint for the community.

\section{Conclusion}
\label{sec:conclusion}
This paper presented a case study on adapting foundation models to specialized, data-scarce industrial domains. Using SEM contour extraction, we proposed \texttt{SegSEM}, a framework proving the viability of a dual-pronged methodology: (1) a data-efficient adaptation strategy that fine-tunes only the encoders of a pre-trained \texttt{SAM2} model, and (2) a robust hybrid architecture that integrates a traditional algorithm as a confidence-aware fallback. Our experiments validated this approach on a small dataset of 60 production images, demonstrating its effectiveness and reliability. 

\small
\bibliographystyle{IEEEtran}
\bibliography{references}

@inproceedings{taberySEMImageContouring2007a,
	title = {SEM Image Contouring for OPC Model Calibration and Verification},
	booktitle = {Advanced Lithography},
	author = {Tabery, Cyrus and Morokuma, Hidetoshi and Matsuoka, Ryoichi and Page, Lorena and Bailey, George E. and Kusnadi, Ir and Do, Thuy},
	year = {2007},
	month = mar,
	pages = {652019},
	address = {San Jose, CA},
	urldate = {2025-04-02},
	langid = {english}
}

@inproceedings{vasekSEMcontourbasedOPCModel2007,
	title = {SEM-Contour-Based OPC Model Calibration through the Process Window},
	booktitle = {Advanced Lithography},
	author = {Vasek, Jim and Menedeva, Ovadya and Levitzky, Dan and Lindman, Ofer and Nemadi, Youval and Bailey, George E. and Sturtevant, John L.},
	year = {2007},
	month = mar,
	pages = {65180D},
	address = {San Jose, CA},
	urldate = {2025-04-02},
	langid = {english}
}

@inproceedings{weisbuchCalibratingEtchModel2015,
	title = {Calibrating Etch Model with SEM Contours},
	booktitle = {SPIE Advanced Lithography},
	author = {Weisbuch, Fran{\c c}ois and Omran, A. and Jantzen, Kenneth},
	year = {2015},
	month = mar,
	pages = {94261T},
	address = {San Jose, California, United States},
	urldate = {2025-03-24},
	langid = {english}
}

@inproceedings{zhengOptimalOPCModel2022,
	title = {Optimal OPC Model Selection with SEM Image Contours},
	booktitle = {2022 International Workshop on Advanced Patterning Solutions (IWAPS)},
	author = {Zheng, Zhen-Fei and Zhang, Wei and Sun, Chen-Wei and Yang, Xiao and Li, Xiao-Mei and Shao, Feng and Zhu, Cynthia and Fenger, Germain and Yu, Yue-Long and Wang, Ying-Fang},
	year = {2022},
	month = oct,
	pages = {1--4},
	publisher = {IEEE},
	address = {Beijing, China},
	urldate = {2025-03-24},
	copyright = {https://doi.org/10.15223/policy-029},
	isbn = {979-8-3503-9766-6},
	langid = {english}
}

@inproceedings{weisbuchAssessingSEMContour2014,
	title = {Assessing SEM Contour Based OPC Models Quality Using Rigorous Simulation},
	booktitle = {SPIE Advanced Lithography},
	author = {Weisbuch, Francois and Samy Naranaya, Aravind},
	year = {2014},
	month = mar,
	pages = {90510A},
	address = {San Jose, California, USA},
	urldate = {2025-03-24},
	langid = {english}
}

@inproceedings{lesireAcceleratingProcessRobustness2024,
 	title = {Accelerating Process Robustness Consolidation with Massive Beam Inspection and CD SEM Contour-Analysis},
 	booktitle = {39th European Mask and Lithography Conference (EMLC 2024)},
 	author = {Lesire, Nicolas and {Le-Gratiet}, Bertrand and Le Pennec, Aur{\'e}lie and Oudin, Louis Victor and Alestra, Romain and Madec, Clementine and Mourier, Lucie and Dettoni, Florent and Deleuze, Pierre Marie and Menis, Audrey},
 	year = {2024},
 	month = sep,
 	pages = {19},
 	publisher = {SPIE},
 	address = {Grenoble, France},
 	urldate = {2025-04-03},
 	isbn = {978-1-5106-8288-7 978-1-5106-8289-4},
 	langid = {english}
 }

@article{kangASFYOLONovelYOLO2024,
	title = {ASF-YOLO: A Novel YOLO Model with Attentional Scale Sequence Fusion for Cell Instance Segmentation},
	shorttitle = {ASF-YOLO},
	author = {Kang, Ming and Ting, Chee-Ming and Ting, Fung Fung and Phan, Rapha{\"e}l C.-W.},
	year = {2024},
	month = jul,
	journal = {Image and Vision Computing},
	volume = {147},
	pages = {105057},
	issn = {02628856},
	urldate = {2025-04-03},
	langid = {english}
}

@inproceedings{gainesScanningElectronMicroscope2024,
	title = {Scanning Electron Microscope Image Segmentation with Foundation AI Vision Model for Nanoparticles in Autonomous Materials Explorations},
	booktitle = {2024 IEEE Conference on Artificial Intelligence (CAI)},
	author = {Gaines, Timothy B. and Boyle, Camden and Keller, James M. and Maschmann, Matthew R. and Price, Stanton and Scott, Grant J.},
	year = {2024},
	month = jun,
	pages = {1266--1271},
	publisher = {IEEE},
	address = {Singapore, Singapore},
	urldate = {2025-04-03},
	copyright = {https://doi.org/10.15223/policy-029},
	isbn = {979-8-3503-5409-6},
	langid = {english}
}

@misc{kousakaAutomatedCellStructure2025,
	title = {Automated Cell Structure Extraction for 3D Electron Microscopy by Deep Learning},
	author = {Kousaka, Jin and Iwane, Atsuko H. and Togashi, Yuichi},
	year = {2025},
	month = feb,
	number = {arXiv:2405.06303},
	eprint = {2405.06303},
	primaryclass = {q-bio},
	publisher = {arXiv},
	urldate = {2025-04-03},
	archiveprefix = {arXiv},
	langid = {english}
}

@article{shiSegmentAnythingModel2025,
	title = {Segment Anything Model for Few-Shot Medical Image Segmentation with Domain Tuning},
	author = {Shi, Weili and Zhang, Penglong and Li, Yuqin and Jiang, Zhengang},
	year = {2025},
	month = jan,
	journal = {Complex \& Intelligent Systems},
	volume = {11},
	number = {1},
	pages = {37},
	issn = {2199-4536, 2198-6053},
	urldate = {2025-04-07},
	langid = {english}
}

@article{liaoApplyingMaskRCNN2024,
	title = {Applying a Mask R-CNN Machine Learning Algorithm for Segmenting Electron Microscope Images of Ceramic Bronze-Casting Moulds},
	author = {Liao, Lingyu and Sun, Zhenfei and Liu, Siran and Ma, Shining and Chen, Kunlong and Liu, Yue and Wang, Yongtian and Song, Weitao},
	year = {2024},
	month = oct,
	journal = {Journal of Archaeological Science},
	volume = {170},
	pages = {106049},
	issn = {03054403},
	urldate = {2025-04-07},
	langid = {english}
}

@article{raviSAM2Segment,
  title={Sam 2: Segment anything in images and videos},
  author={Ravi, Nikhila and Gabeur, Valentin and Hu, Yuan-Ting and Hu, Ronghang and Ryali, Chaitanya and Ma, Tengyu and Khedr, Haitham and R{\"a}dle, Roman and Rolland, Chloe and Gustafson, Laura and others},
  journal={arXiv preprint arXiv:2408.00714},
  year={2024}
}

@inproceedings{zhouEvaluationHotspotsEPE2024,
	title = {Evaluation of Hotspots EPE Propagation through Step-by-Step SEM Contour Analysis},
	booktitle = {Eighth International Workshop on Advanced Patterning Solutions (IWAPS 2024)},
	author = {Zhou, Kan and Zhou, Wenzhan and Meng, Yuanyuan and Su, Xiaohang and Wang, Libo and Li, Xiaoci and Liu, Zhengfang and Du, Chunshan and Dou, Huaiyang},
	year = {2024},
	month = dec,
	pages = {48},
	publisher = {SPIE},
	address = {Jiaxing, China},
	urldate = {2025-04-14},
	isbn = {978-1-5106-8632-8 978-1-5106-8633-5},
	langid = {english}
}

@inproceedings{raniEdgeDetectionScanning2020,
	title = {Edge Detection in Scanning Electron Microscope (SEM) Images Using Various Algorithms},
	booktitle = {2020 4th International Conference on Intelligent Computing and Control Systems (ICICCS)},
	author = {Rani, G Elizabeth and Murugeswari, R and Rajini, N},
	year = {2020},
	month = may,
	pages = {401--405},
	publisher = {IEEE},
	address = {Madurai, India},
	urldate = {2025-04-14},
	copyright = {https://ieeexplore.ieee.org/Xplorehelp/downloads/license-information/IEEE.html},
	isbn = {978-1-7281-4876-2},
	langid = {english}
}

@inproceedings{heCDSEMContourExtraction2022,
	title = {CD-SEM Contour Extraction for Complex Features Measurement},
	booktitle = {2022 China Semiconductor Technology International Conference (CSTIC)},
	author = {He, Ting and Zhang, Yingchun and Wang, Jian},
	year = {2022},
	month = jun,
	pages = {1--5},
	publisher = {IEEE},
	address = {Shanghai, China},
	urldate = {2025-03-24},
	copyright = {https://doi.org/10.15223/policy-029},
	isbn = {978-1-6654-9758-9},
	langid = {english}
}

@article{zhangMulticomponentSegmentationXray2017,
	title = {Multi-Component Segmentation of X-Ray Computed Tomography (CT) Image Using Multi-Otsu Thresholding Algorithm and Scanning Electron Microscopy},
	author = {Zhang, Pengfei and Lu, Shuangfang and Li, Junqian and Zhang, Ping and Xie, Liujuan and Xue, Haitao and Zhang, Jie},
	year = {2017},
	month = may,
	journal = {Energy Exploration \& Exploitation},
	volume = {35},
	number = {3},
	pages = {281--294},
	issn = {0144-5987, 2048-4054},
	urldate = {2025-04-25},
	langid = {english}
}

@article{zhao2024adaptive,
  title={Adaptive thresholding and coordinate attention-based tree-inspired network for aero-engine bearing health monitoring under strong noise},
  author={Zhao, Dezun and Cai, Wenbin and Cui, Lingli},
  journal={Advanced Engineering Informatics},
  volume={61},
  pages={102559},
  year={2024},
  publisher={Elsevier}
}

@article{pang2024suppression,
  title={Suppression of noise in SEM images using adaptive anisotropic partial differential equations},
  author={Pang, Shuiquan and Zhang, Qiuzhen and Wang, Zhizhe and Luo, Jun and Wu, Shihuang and Li, Hai and Zhang, Xianmin},
  journal={Journal of Micro and Bio Robotics},
  volume={20},
  number={2},
  pages={1--13},
  year={2024},
  publisher={Springer}
}

@article{sauvola2000adaptive,
  title={Adaptive document image binarization},
  author={Sauvola, J. and Pietik{\"a}inen, M.},
  journal={Pattern recognition},
  volume={33},
  number={2},
  pages={225--236},
  year={2000},
  publisher={Elsevier}
}

\end{document}